\documentstyle[12pt]{article}

\newcommand{\nn}{\nonumber\\}\newcommand{\p}[1]{(\ref{#1})}
\begin{document}
\renewcommand{\thefootnote}{\fnsymbol{footnote}}
\thispagestyle{empty}
{\hfill  Preprint JINR E2-97-122}\vspace{0.5cm} \\

{\hfill \bf Dedicated to the memory of$\;\;$}\vspace{-0.5cm}\\

{\hfill \bf D.V. Volkov and V.I. Ogievetsky}\vspace{2.5cm} \\
\begin{center}
{\large\bf Nonlinear realizations of the (super)diffeomorphism groups,
geometrical objects and integral invariants in the superspace.}
 \vspace{1.5cm} \\
A. Pashnev\footnote{BITNET: PASHNEV@THSUN1.JINR.DUBNA.SU}\vspace{1cm}\\
{\it JINR--Bogoliubov Laboratory of Theoretical Physics} \\
{\it Dubna, Head Post Office, P.O.Box 79, 101 000 Moscow, Russia}
\vspace{1.5cm} \\
{\bf Abstract}
\end{center}
It is shown that vielbeins and connections of any (super)space are 
naturally described in terms of nonlinear realizations of
infinite - dimensional diffeomorphism groups of the corresponding
(super)space. The method of construction of integral invariants
from the invariant Cartan's differential $\Omega$ - forms is generalized
to the case of superspace.
\begin{center}
{\it Submitted to ``Classical and Quantum Gravity"}
\end{center}
\vfill
\setcounter{page}0
\renewcommand{\thefootnote}{\arabic{footnote}}
\setcounter{footnote}0
\newpage
\section{Introduction}
As was shown in \cite{BO}, gravity can be realized as a nonlinear
realization of the four dimensional diffeomorphism group.
The consideration was based on the fact that infinite dimensional
diffeomorphism group in four dimensional space can be represented as the
closure of two finite dimensional groups - conformal and affine \cite{OP}.
As a consequence of such representation of the diffeomorphism group,
the basic field in this consideration was the symmetric tensor
field of the second rank - the metric field $g_{m,n}$, which 
corresponds to symmetric generators of the affine group.

The generalization of this approach to the case of superspace was
given in \cite{IN}.

In the present work we consider nonlinear realization of the whole
infinite dimensional diffeomorphism group of the arbitrary 
(super)space. Among the coordinates parametrizing the group element
(coset space)
in such realization there are present
usual coordinates of the (super)space.
The (super)vielbein and (super)connection are naturally represented
as the functions of other coordinates of the coset space.
The structure of the connection
in the purely bosonic case is such that the corresponding torsion
is zero. In the superspace only some components of the torsion,
namely $T^a_{bc}$ and $T^\alpha_{\beta\gamma}$, vanish automatically.

\setcounter{equation}0\section{Bosonic space}
Firstly we consider the case of the usual $D$ - dimensional bosonic space
with the coordinates $s^m,\; m=0,1,...,D-1$. The generators of the
corresponding diffeomorphism group regular 
at the origin can be written in the coordinate
representation as
\begin{equation}                \label{rep}
{P^{m_1,m_2,...,m_n}}_m=is^{m_1}s^{m_2}... s^{m_n}\frac{\partial}
{\partial s^m} .
\end{equation}

All of this generators can be naturally ordered in accordance with
their dimensionality ($dim \;s^m=+1$):
\begin{equation}
dim\;P_m=-1,\;dim\;{P^{m_1}}_m=0,\;dim\;{P^{m_1,m_2}}_m=+1,\;
dim\;{P^{m_1,m_2,m_3}}_m=+2,...\; .
\end{equation}
With the help of the representation \p{rep} one can 
calculate the commutation
relations between the generators of the diffeomorphism group and 
after that we can forget about the auxiliary coordinates $s^m$. The
only we will need is the following algebra:
\begin{eqnarray}\label{algebra}
&&[{P^{m_1,m_2,...,m_n}}_m\; ,
{P^{k_1,k_2,...,k_l}}_k]\;=\\
&&i\sum_{i=1}^{l}\delta_m^{k_i}{P^{m_1,...m_n,k_1...
k_{i-1}k_{i+1}...,k_l}}_k
-i\sum_{j=1}^{n}\delta_k^{m_j}{P^{m_1,...m_{j-1}m_
{j+1}...m_n,k_1...,k_l}}_m. \nonumber
\end{eqnarray}

Let us consider  the following parametrization of the group element:
\begin{eqnarray}\label{coset}
&&G=e^{ix^mP_m}e^{i{\phi^l}_{l_1,l_2}{P^{l_1,l_2}}_l}
e^{i{\phi^k}_{k_1,k_2,k_3}{P^{k_1,k_2,k_3}}_k}
 ... e^{i{\phi^n}_{n1}P^{n_1}_n}.
\end{eqnarray}
All parameters in the expression \p{coset} are symmetric with respect to
the permutation of lower indices as a consequence of the symmetry 
properties of the generators \p{rep}.
It is convenient  to take the element of the finite dimensional group
$GL(D)$, generated by $P^{n_1}_n$, as the last multiplier in the
expression \p{coset}. The rest of the factors in \p{coset} are
ordered with respect to the dimensionality of the generators. 
As a consequence, the product of factors to the right from arbitrary
one form a subgroup of the diffeomorphism group.
Such structure of the group element simplifies the evaluation of the 
variations $\delta {\phi^l}_{l_1,...,l_n}$ under the infinitesimal
left action
\begin{equation}\label{left}
G'=(1+i\epsilon)G,
\end{equation}
where $\epsilon =
\epsilon^m P_m+{\epsilon^{m}}_{m_1}{P^{m_1}}_m+
{\epsilon^{m}}_{m_1,m_2}{P^{m_1,m_2}}_m+... $ belongs to the
algebra of the diffeomorphism group.
The coordinates in \p{coset} transform through the infinitesimal
transformation parameters ${\epsilon^m}_{m_1,m_2,...,m_k}$ 
and coordinates wich are placed to the left from given ones in
parametrization \p{coset}:
\begin{eqnarray}
&&\delta x^m=\delta x^m(\epsilon ,x^i),\;
\delta {\phi^n}_{n_1}=\delta {\phi^n}_{n_1}(\epsilon ,x^m,
{\phi^k}_{k_1}),\\
&&\delta {\phi^l}_{l_1,l_2}=
\delta {\phi^l}_{l_1,l_2}(\epsilon,\; x^i,\;
{\phi^k}_{k_1,k_2}),... .
\end{eqnarray}
The only exception is the transformation law for $\phi^n_{n_1}$,
which includes only $\epsilon , x^m$ and $\phi^n_{n_1}$ itself.
At this stage it is natural to consider all parameters as the fields
in $D$ - dimensional space parametrized by coordinates $x^m$.

Step by step one can evaluate the variations of all parameters of the
coset. The general method of calculations is as follows.
To find the variation $\delta {\phi^l}_{l_1,...,l_n}$ we have to
solve the equation
\begin{equation}\label{1}
(1+i\epsilon)e^{i\phi^n}=e^{i\phi^n+i\delta \phi^n}(1+i\tilde\epsilon).
\end{equation}
where, for the brevity, $\phi^n={\phi^l}_{l_1,...,l_n}{P^{l_1,...,l_n}}_l$
and parameter $\epsilon$ contains the generators
with $n$ and more upper indices. Correspondingly, $\tilde\epsilon$
contains the generators with $n+1$ or more upper indices. Both of these
parameters contain $P_n^m$.

From \p{1} it simply follows:
\begin{equation}
ie^{-i\phi^n}\epsilon e^{i\phi^n}=e^{-i\phi^n}\delta e^{i\phi^n}+
i\tilde\epsilon.
\end{equation}
Both right and left part of this equation can be written in terms
of multiple commutators
\begin{equation} \label{2}
e^{-i\phi^n} \wedge \epsilon=\frac{e^{-i\phi^n}-1}{i\phi^n}
 \wedge \delta \phi^n + \tilde\epsilon,
\end{equation}
where, for simplicity, we use the notation
\begin{equation}
e^{-i\phi^n} \wedge \epsilon=\epsilon+\frac{1}{1 !}[-i\phi^n,\epsilon]+
\frac{1}{2 !}[-i\phi^n,[-i\phi^n,\epsilon]]+... .\nonumber
\end{equation}
The equation \p{2} is the basic equation for 
$\delta\phi^n$ and $\tilde\epsilon$.

The simplest transformation law have the dimension-one coordinates
$x^m$. They transform as the coordinates 
of the $D$ -dimensional space under the reparametrization:
\begin{equation}\label{x}
\delta x^m =\varepsilon^m(x)\equiv\epsilon^m+\epsilon^m_{m_1}x^{m_1}+
\epsilon^m_{m_1m_2}x^{m_1}x^{m_2}+...\;.
\end{equation}
Here $\varepsilon^m(x)$ is infinitesimal function of the
coordinates $x^n$.
This is a consequence of first among the relations \p{1}:
\begin{equation}\label{1}
(1+i\epsilon)e^{ix^mP_m}=e^{i(x^m+\delta x^m)P_m}(1+i\tilde\epsilon),
\end{equation}
in which $\delta x^m$ is given by \p{x} and:
\begin{equation}
\tilde\epsilon=\frac{1}{1!}\partial_{m_1}\epsilon^{m}{P^{m_1}}_m+
\frac{1}{2!}\partial_{m_1m_2}\epsilon^{m}{P^{m_1,m_2}}_m+... \; .
\end{equation}

The next parameters in the coset have three indices and transform as a
Cristoffel symbol:
\begin{eqnarray}
&&\delta {\phi^m}_{m_1m_2}=
\frac{\partial \varepsilon^m}{\partial x^n}{\phi^n}_{m_1m_2}
-\frac{\partial \varepsilon^n}{\partial x^{m_2}}{\phi^m}_{m_1n}-
\frac{\partial \varepsilon^n}{\partial x^{m_1}}{\phi^m}_{nm_2}+
\frac{1}{2}\frac{\partial^2 \varepsilon^m}
{\partial x^{m_1}\partial x^{m_2}} .
\end{eqnarray}
In general the transformation law for parameter with $n$ lower indices
will contain the term with $n$-th derivative of infinitesimal 
parameter $\epsilon^m(x)$.

Only variations of the last parameters ${\phi^n}_{n_1}$ need the
separate consideration. To find them one have to evaluate the
expression
\begin{equation}
\delta (e^{i{\phi^n}_{n1}P^{n_1}_n}) =i
\frac{\partial \varepsilon^m(x)}{\partial x^k}
P^k_m\; e^{i{\phi^n}_{n1}P^{n_1}_n}
\end{equation}
The simplest way to do this is to use the matrix representation for
the generators of $GL(D)$ group:
\begin{equation}
(P^{n_1}_n)_k^l=i\delta_k^{n_1}\delta_n^l.
\end{equation}
In this representation the element of $GL(D)$ group is the exponent of 
the matrix $\phi_n^{m}$:
\begin{equation}\label{Ekl}
(e^{i{\phi^n}_{n_1}P^{n_1}_n})_k^l=(e^{-\phi})_k^l\equiv E_k^l.
\end{equation}
It is convenient to consider the matrix $E_k^l$ instead of $\phi_n^m$
because its transformation law is very simple:
\begin{equation}
\delta E_k^l=-\frac{\partial \varepsilon^m(x)}{\partial x^k}E_m^l.
\end{equation}
It means that the $E_k^l$ transforms as the covariant vector with respect
to its lower index. Simultaneously, its upper index is inert. This is the
transformation law of the vielbein.

The fact that $E_k^l$ is endeed the vielbein becomes evident if we
consider the Cartan's differential form
\begin{equation}
\Omega=G^{-1}dG=i\Omega^aP_a+i\Omega_a^bP_b^a+i\Omega_{a_1a_2}^b
P_b^{a_1a_2}+... ,
\end{equation}
which simultaneously with its components $(\Omega^a,\;\Omega^a_b,...)$
is invariant with respect to the left transformation \p{left}.
We emphasize the fact of invariance by using the letters from the
beginning of the alphabet for indices.
The explicit expressions for the components of the $\Omega$ -form are:
\begin{eqnarray}\label{a}
\Omega^a&=&E^a_mdx^m,\\  \label{ab}
\Omega_b^a&=&-E_b^mdE_m^a-2dx^k\phi^m_{kn}E_b^nE_m^a,\\ \label{abc}
\Omega^a_{bc}&=&(d\phi^m_{kn}-dx^l\phi^m_{il}\phi^i_{kn}+
dx^l\phi^m_{in}\phi^i_{kl}+dx^l\phi^m_{ik}\phi^i_{ln}-
3dx^l\phi^m_{knl})E^a_mE^k_bE^n_c,...\;.
\end{eqnarray}
The first of this forms is exactly one-form vielbein. The physical
meaning of its index $a$ becomes clear if we consider the right gauge
transformation belonging to $GL(D)$
\begin{equation}\label{gauge}
G'=G\{1-ih(x)\}=G\{1-ih_b^a(x)P_a^b\}
\end{equation}
Vielbein one -form $E^a\equiv\Omega^a$ transforms as the vector
\begin{equation}
\delta E^a=h_b^a E^b
\end{equation}
of this $GL(D)$, which can be
considered as the gauge group in the tangent space. 
All $\Omega$ -forms with higher number of indices transform homogeneously
as corresponding tensors.
The only exception is the differential one-form \p{ab} which transforms
inhomogeneously:
\begin{equation}
\delta\Omega^a_b=h^a_c\Omega^c_b-\Omega^a_ch^c_b-dh^a_b.
\end{equation}
This is exactly the transformation law of the connection one-form and
$\Omega^a_b$ is the natural candidate for the connection in the absence
of any other tensors of second rank, which could be, in principle,
added to the connection. So, the ``minimal" 
one-form connection is given by
\p{ab} in terms of vielbein $E_m^a$ and Cristoffel symbol $\phi^m_{kn}$.
The corresponding curvature two -form
\begin{equation}         \label{rab}
R^a_b=d\Omega^a_b+\Omega^a_c\Omega^c_b
\end{equation}
transforms as a tensor of second rank.

Due to its definition, the $\Omega$ -form satisfy the Maurer-Cartan
equation
\begin{equation}
d\Omega+\Omega\wedge\Omega=0,
\end{equation}
or, in components
\begin{eqnarray}\label{da}
d\Omega^a&+&\Omega^a_b\Omega^b=0,\\ \label{dab}
d\Omega^a_b&+&\Omega^a_c\Omega^c_b+2\Omega^a_{bc}\Omega^c=0,\\ \label{dac}
d\Omega^a_{bc}&+&\Omega^a_d\Omega^d_{bc}+\Omega^a_{bd}\Omega^d_c+
\Omega^a_{cd}\Omega^d_b+3\Omega^a_{bcd}\Omega^d=0,...\; .
\end{eqnarray}
The lefthand side of first of these equations \p{da} is the covariant
differential of the vielbein with the connection $\Omega^a_b$. The fact
of its equality to zero means vanishing of the corresponding torsion.
Second equation represents the curvature two-form \p{rab}
in terms of vielbein and
$\Omega$-form with three indices
\begin{equation}
R^a_b=-2\Omega^a_{bc}\Omega^c.
\end{equation}
The rest of equations express covariant differentials of $\Omega$-forms
in terms of other $\Omega$-forms.

The following expression for the action
\begin{equation} \label{act}
S=\int R^{a_1}_b \eta^{b a_2}\Omega^{a_3}...\Omega^{a_D}
\varepsilon_{a_1a_2...a_D}
\end{equation}
leads to Einstein - Gilbert action 
\begin{equation}
S=\int d^Dx \sqrt{-g}R,\;\;\;\;\; g^{mn}=\eta^{ab}E_a^mE_b^n,
\end{equation}
after elimination of $\phi^m_{kn}$ 
with the help of its equation of motion
in terms of $g^{mn}$.
Some comments are needed here. Up to now the gauge group in tangent
space, considered as right transformations \p{gauge}, was group
$GL(D)$ of general linear transformations in\\ $D$ - dimensions.
In principle, one can construct action, invariant under the whole
$GL(D)$ gauge group, for example $\int R_a^b R_b^a$ in four
dimensions.
The presence in the action \p{act} of two constant tensors - absolutely
antisymmetric tensor $\varepsilon_{a_1a_2...a_D}$ and tangent
space flat metric $\eta^{ab}=diag(1,1,...,1,-1)$ means that
the invariance group of the action \p{act} is the subgroup of $GL(D)$,
namely, the group $SO(D-1,1)$. So, the choice of the gauge group
in the tangent space depends on the structure constants in the action,
which can break $GL(D)$ down to its subgroup.

\setcounter{equation}0\section{Superspace}
As a generalization of the approach we consider the diffeomorphism
group of the superspace with coordinates $s^M$, from which $D$ 
coordinates 
$s^m,\; m=0,1,...,D-1$ are bosonic  and $D_G $ coordinates 
$\eta^\mu,\;\mu=1,2,...,D_G$ are grassmann. Both numbers $D$ and $D_G$
from the very beginning are arbitrary. The grassmann grading of the
coordinates $g(s^m)=0,\;g(s^\mu)=1$ means the standard commutation
relations: $s^Ms^N-(-1)^{g(s^M)g(s^N)}s^Ns^M=0$, or, for the brevity,
$s^Ms^N-(-1)^{MN}s^Ns^M=0$. The generators of the algebra
\begin{equation}                \label{reps}
{P^{M_1,M_2,...,M_n}}_M=is^{M_1}s^{M_2}... s^{M_n}\frac{\partial}
{\partial s^M} .
\end{equation}
have the following dimensionalities:
\begin{eqnarray}
&&dim\;P_m=-1,\;dim\;P_\mu=dim\;{P^{\mu}}_m=-\frac{1}{2} ,\;\nn       
&&dim\;{P^{m_1}}_m=dim\;{P^{\mu_1}}_\mu=dim\;{P^{\mu_1\mu_2}}_m=0,\;\\
&&dim\;{P^{\mu_1\mu_2}}_\mu=dim\;{P^{m_1\mu}}_m=+\frac{1}{2} ,\;
dim\;{P^{m,\mu_1}}_\mu=dim\;{P^{m_1m_2}}_m=+1,...\; .\nonumber
\end{eqnarray}
Some of the generators are bosonic and others (with halfinteger
dimensionality) - fermionic with grassmann grading $0$ or $1$ 
correspondingly. The same grading $0$ or $1$ corresponds to bosonic $m$
or fermionic $\mu$ indices. The algebra of the generators \p{reps}
is graded algebra:
\begin{eqnarray}
&&{P^{M_1,M_2,...,M_n}}_M{P^{N_1,N_2,...,N_k}}_N-
(-1)^{(M_1+...+M_n+M)(N_1+...+N_k+N)}
{P^{N_1,...,N_k}}_N{P^{M_1,...,M_n}}_M=\nn
&&i\sum_{l=1}^{k} {\delta_M}^N_l(-1)^{M(N_1+...+N_{l-1})}
{P^{M_1...M_nN_1...N_{l-1}N_{l+1}...N_k}}_N-\\
&&-i\sum_{l=1}^{n}{\delta_N}^{M_l}
(-1)^{(M_1+...+M_n+M)(N_1+...+N_k+N)+N(M_1+...+M_{l-1})}
{P^{N_1...N_kM_1...M_{l-1}M_{l+1}...M_n}}_M.\nonumber
\end{eqnarray}

It is convenient to parametrize the group element in the form
\begin{equation}
G=K H,
\end{equation}
where
\begin{eqnarray}
K&=&e^{ix^mP_m}e^{i\theta^\mu P_{\mu}}e^{i{u^M}_{M_1M_2}{P^{M_2M_1}}_M}
e^{i{u^N}_{N_1N_2N_3}{P^{N_3N_2N_1}}_N}... \; ,\\
&&\nn \label{KS}
H&=&e^{i{\psi^m}_\mu{P^\mu}_m}e^{i{\phi^\nu}_n{P^n}_\nu}
e^{i{u^k}_l{P^l}_k}e^{i{v^\rho}_\sigma{P^\sigma}_\rho}.
\end{eqnarray}
The element $H$ belongs to the finite - dimensional subgroup
$GL(D,D_G)$ of the superdiffeomorphism group and its parameters have
dimensions: $dim\;{\psi^m}_\mu=1/2, \; dim\;{\phi^\nu}_n=-1/2,\;
dim\;{u^k}_l=dim\;{v^\rho}_\sigma=0$. The coset $K=G/H$ is
parametrized by infinite number of the parameters with dimensions:
$dim\;x^m=1,\;dim\;\theta^\mu=1/2$, $\;\;dim\;{u^N}_{N_1N_2}$ run
from $0$ to $-3/2$ etc.

Consider the element of the diffeomorphism algebra
\begin{equation}
\epsilon =\epsilon^M P_M+{\epsilon^{M}}_{M_1}{P^{M_1}}_M+
{\epsilon^{M}}_{M_1M_2}{P^{M_2M_1}}_M+... .
\end{equation}
with the constant infinitesimal coefficients.
Under the left action
\begin{equation}\label{lefts}
G'=(1+i\epsilon)G,
\end{equation}
the parameters $x^M$ = $(x^m,\; \theta^{\mu})$ transform as the
coordinates of the $(D,D_G)$ - dimensional superspace:
$\delta x^M =\varepsilon^M(x),$
where 
\begin{equation}
\varepsilon^M(x)=\epsilon^M+{\epsilon^M}_{M_1}x^{M_1}+
{\epsilon^M}_{M_1M_2}x^{M_2}x^{M_1}+...\;.
\end{equation}
The rest of the parameters in \p{KS} transform in a more complicated way.
Exactly as in the bosonic case the transformation laws of the 
parameters with $n$ lower
indices includes all parameters up to $n$ lower indices and all
derivatives of $\varepsilon^M(x)$ up to $n$ -th order. The calculation 
of this transformation laws is complitely analogous to the purely bosonic
case.

The next to the right after $x^M$ are parameters with three indices:
${u^M}_{M_1M_2}$.
They transform inhomogeneously as the Cristoffel symbols in the
superspace with coordinates $x^M$
\begin{eqnarray}
&&\delta {u^M}_{M_1M_2}=
(-1)^{N(M+1)}\frac{\partial}{\partial x^N}\varepsilon^M{u^N}_{M_1M_2}
-(-1)^{M_2(N+1)}{u^M}_{M_1N}\frac{\partial}
{\partial x^{M_2}}\varepsilon^N-\nn
&&{}\\
&&(-1)^{(M_1+M_2)(M_1+N)}{u^M}_{NM_2}\frac{\partial}
{\partial x^{M_1}}\varepsilon^N+
\frac{1}{2}(-1)^{(M+1)(M_1+M_2)}
\frac{\partial}{\partial x^{M_1}}  
\frac{\partial}{\partial x^{M_2}}\varepsilon^M .\nonumber 
\end{eqnarray}

The transformation laws of the components of the supervielbein are
as follows:
\begin{eqnarray}\label{sv1}
\delta{\psi^m}_\mu&=&-\partial_\mu\epsilon^m+\partial_n\epsilon^m
{\psi^n}_\mu-\partial_\mu\epsilon^\nu{\psi^m}_\nu-
\partial_n\epsilon^\nu{\psi^n}_\mu{\psi^m}_\nu=\\
&&-D_\mu\epsilon^m-D_\mu\epsilon^\nu{\psi^m}_\nu,\;\;
D_\mu=\partial_\mu-{\psi^m}_\mu\partial_m,\nn
\delta{\phi^\mu}_m&=&\partial_m\epsilon^\mu-(\partial_m\epsilon^n+
\partial_m\epsilon^\nu{\psi^n}_\nu){\phi^\mu}_n+
+(\partial_\nu\epsilon^\mu+
\partial_n\epsilon^\mu{\psi^n}_\nu){\phi^\nu}_m,\\
\delta{{\cal E}_k}^a&=&-(\partial_k\epsilon^m+
\partial_k\epsilon^\nu{\psi^m}_\nu){{\cal E}_m}^a,\\ \label{sv4}
\delta{{\cal E}_\mu}^\alpha&=&-(\partial_\mu\epsilon^\rho+
\partial_m\epsilon^\rho{\psi^m}_\nu){{\cal E}_\rho}^\alpha=
-D_\mu\epsilon^\rho{{\cal E}_\rho}^\alpha.
\end{eqnarray}
In analogy with \p{Ekl} we denote ${\cal E}^a_m=(e^{-u})^a_m,
\;\;{\cal E}^\alpha_\mu=(e^{-v})^\alpha_\mu$.
The next step is to consider all parameters as the fields in the
superspace with $D$ bosonic and $D_G$ grassmann coordinates $x^M$
and construct invariant differential forms in terms of these fields.

\setcounter{equation}0\section{Differential $\Omega$ - forms in the
superspace}

Along with grading of the coordinates $x^M$, their differentials $dx^M$
have their own grading. There exist two different gradings of the 
differentials of the coordinates. One of them corresponds to the
independent grassmann grading and grading of the differential $d$\cite{W}.
It leads to the following commutation relations:
\begin{eqnarray}
[x^m,x^n]&=&[x^m,dx^n]=[x^m,d\theta^\mu]=[d\theta^\mu,d\theta^\nu]=
[\theta^\mu,dx^m]=0,\nn
\{dx^m,dx^n\}&=&\{\theta^\mu,\theta^\nu\}=\{dx^m,d\theta^\mu\}=
\{\theta^\mu,d\theta^\nu\}=0.
\end{eqnarray}
More simple commutation relations take place when grading of the
differential $d$ coincides with grassmann grading \cite{B}. It means that 
the differential changes the grading of the coordinates to the
opposite one:
 \begin{equation}\label{g}
g(dx^M)=g(x^M)+1.
\end{equation}  
 As a result there are equal
numbers $D+D_G$ of bosonic $(x^m,\;d\theta^\mu)$ and grassmann
$(dx^m,\;\theta^\mu)$ variables. 

The left - invariant differential $\Omega$ -form 
\begin{equation}\label{omegas}
\Omega=G^{-1}dG
\end{equation}
belongs to the algebra of the superdiffeomorphism group
\begin{equation}
\Omega=i\Omega^AP_A+i{\Omega^A}_{A_1}{P^{A_1}}_A+
i{\Omega^A}_{A_1A_2}{P^{A_2A_1}}_A+... 
\end{equation}
and its coefficients $\Omega^A,\;{\Omega^A}_{A_1},\;{\Omega^A}_{A_1A_2}$
are invariant under the transformation \p{lefts}. 
We emphasize this
fact by using the letters from the beginning of the alphabet for
indices.
We will use latin letters $a,b,c,...$ for bosonic and
greek letters $\alpha,\beta,\gamma,...$ for grassmann indices.
Note, that according to the grading rule \p{g} $\Omega_a$ and 
$\Omega^\alpha$ are, correspondingly, anticommuting and commuting
objects.

Explicit expressions for components of $\Omega^A$ are:
\begin{eqnarray} \label{oa}
\Omega^a&=&(dx^m+d\theta^\mu{\psi^m}_\mu){{\cal E}_m}^a\equiv
dx^M{E_M}^a,\\ \label{oal}
\Omega^\alpha&=&\{d\theta^\mu-
(dx^m+d\theta^\nu{\psi^m}_\nu){\phi^\mu}_m\}{{\cal E}_\mu}^\alpha\equiv
dx^M{E_M}^\alpha.
\end{eqnarray}
The expresions \p{oa} and \p{oal} represent the one - form supervielbein
$E^A\equiv\Omega^A=dx^M{E_M}^A$ with components\\
\begin{equation}\label{svb}
{}
\end{equation}\vspace{-1.5cm}\\
\begin{tabular}{c|ll|r}
  & ${E_m}^a={{\cal E}_m}^a$ & ${E_m}^\alpha=-{\phi^\mu}_m
{{\cal E}_\mu}^\alpha$&\\
${E_M}^A=$&&&\\
  & ${E_\mu}^a={{\cal E}_m}^a{\psi^m}_\mu$ & 
${E_\mu}^\alpha={{\cal E}_\mu}^\alpha+{{\cal E}_\nu}^\alpha
{\phi^\nu}_n{\psi^m}_\mu$&\\
\end{tabular}\vspace{0.3cm}\\
From \p{sv1}-\p{sv4} it follows
\begin{equation}
\delta{E_M}^A=-\epsilon^N\partial_N{E_M}^A-\partial_M\epsilon^N{E_N}^A.
\end{equation}
The components  of the inverse supervielbein are\\
\begin{equation}\label{isvb}
{}
\end{equation}\vspace{-1.5cm}\\
\begin{tabular}{c|ll|}
  & ${E_a}^m={{\cal E}_a}^m-{{\cal E}_a}^n{\phi^\nu}_n{\psi^m}_\nu$ &
 ${E_a}^\mu={\phi^\mu}_n{{\cal E}_a}^n$\\
${E_A}^M=$&&\\
  & ${E_\alpha}^m=-{{\cal E}_\alpha}^\nu{\psi^m}_\nu$ & 
${E_\alpha}^\mu={{\cal E}_\alpha}^\mu$\\
\end{tabular}\vspace{0.3cm}\\
where ${{\cal E}_a}^m$  and ${{\cal E}_\alpha}^\mu$ are inverse matrices to
${{\cal E}_m}^a$  and ${{\cal E}_\mu}^\alpha$ correspondingly.
Straightforward computation shows very simple form of the
superdeterminant $Ber E_M^A$
\begin{equation}\label{ber}
Ber E_M^A=\frac{det{{\cal E}_m}^a}{det {{\cal E}_\mu}^\alpha}. 
\end{equation}
One can show that arbitrary nonsingular graded matrix $E_M^A$ can be
parametrized in the form \p{svb} for which \p{ber} is valid. So,
in some sense such parametrization of graded matrices is natural.

Due to its definition \p{omegas} $\Omega$ -form satisfy the
Maurer - Cartan equation 
\begin{equation}
d\Omega+\Omega\wedge\Omega=0.
\end{equation}
Two first components of this equation are as follows:
\begin{eqnarray} \label{dif}
&&d\Omega^A+(-1)^{A+B}{\Omega^A}_B\Omega^B,\\
&&d{\Omega^A}_B+(-1)^{A+C}{\Omega^A}_C{\Omega^C}_B+
2(-1)^{A+B+C}{\Omega^A}_{BC}\Omega^C.
\end{eqnarray}
Under the right gauge transformations from the group 
$GL(D)\times GL(D_G)$:
\begin{equation}
G'=G\{1-ih_b^a(x)P_a^b-ih_\beta^\alpha(x)P_\alpha^\beta\}
\end{equation}
$\Omega^a_b$ and $\Omega^\alpha_\beta$ transform as corresponding
connections:
\begin{eqnarray}
\delta\Omega^a_b&=&h^a_c\Omega^c_b-\Omega^a_ch^c_b-dh^a_b\\
\delta\Omega^\alpha_\beta&=&h^\alpha_\gamma\Omega^\gamma_\beta
-\Omega^\alpha_\gamma h^\gamma_\beta-dh^\alpha_\beta.
\end{eqnarray}
Taking them as a ``minimal" connectiions,
the equation \p{dif} expresses the covariant
differentials of $\Omega^a$ and $\Omega^\alpha$:
\begin{eqnarray}
D\Omega^a&\equiv&d\Omega^a+\Omega^a_b\Omega^b=\Omega^a_\alpha\Omega^\alpha=
T^a_{B\alpha}\Omega^B\Omega^\alpha,\\
D\Omega^\alpha&\equiv&d\Omega^\alpha+\Omega^\alpha_\beta
\Omega^\beta=\Omega^\alpha_a\Omega^a=
T^\alpha_{Ba}\Omega^B\Omega^a,
\end{eqnarray}
where we expanded one forms $\Omega^a_\alpha$ and $\Omega^\alpha_a$
in terms of the basic system of one-forms $\Omega^B$. The form of the
right hand sides of these equations shows, that $T^a_{bc}$ and
$T^\alpha_{\beta\gamma}$ components of the torsion vanish identically.

\setcounter{equation}0\section{Integral invariants in the superspace}
Having constructed invariant differential $\Omega$ - forms we have to be
able to build from them the integral invariants like action. The
problem lies in the fact that there are two types of differentials
of grassmann coordinates. We will denote them $d\theta^\mu$ and
$\underline{d}\theta^\mu$. First of these differentials are used in
invariant differential $\Omega$ - forms.
Unlike $dx^m$, which are anticommuting objects,
$d\theta^\mu$ commute and one can construct from the $\Omega$ -forms
the differential form of arbitrary order.
Contrarily,  so called Berezin pseudodifferentials $\underline{d}\theta^\mu$
anticommute. One can take integrals over the superspace
\begin{equation}\label{int}
I=\int d^Dx\underline{d}^{D_G}\theta F(x,\theta) 
\end{equation}
with the help of Berezin integration rules
\begin{equation}
\int\underline{d}\theta^\mu=0,\;\int\theta^\mu\underline{d}\theta^\nu=
\delta^{\mu\nu}.
\end{equation}
The integration volume $dv=d^Dx\underline{d}^{D_G}\theta$ transforms
under the general diffeomorphism $x'^M=x'^M(x^N)$ as
\begin{equation}
dv'=Ber\left(\frac{\partial x'^M}{\partial x^N}\right)dv. 
\end{equation}
If the function $F(x,\theta)$ in \p{int} transforms as
\begin{equation}
F(x',\theta')=Ber^{-1}\left(\frac{\partial x'^M}{\partial x^N}\right)
F(x,\theta),
\end{equation}
the integral $I$ is reparametrization invariant.

The following prescription for building of integral invariants from 
invariant differential forms was formulated by Bernstein and Leites
\cite{BL}. Taking arbitrary invariant 
differential form $F(x,\theta,dx,d\theta)$
one can introduce new auxiliary variables 
$\eta^m$ and $y^\mu$ in one to one
correspondence to $dx^m$ and $d\theta^\mu$ .These new variables $\eta^M$
have the same grassmann properties as $dx^m$ and $d\theta^\mu$ (opposite
to ones of $x^m$ and $\theta^\mu$) and the
following transformation laws
\begin{eqnarray}\label{auxtr}
\eta'^m&=&\frac{\partial x'^m}{\partial x^n}\eta^n+
\frac{\partial x'^m}{\partial \theta^\mu}y^\mu,\\\label{auxtr1}
y'^\mu&=&\frac{\partial \theta'^\mu}{\partial x^m}\eta^m+
\frac{\partial \theta^\mu}{\partial \theta^\nu}y^\nu.  
\end{eqnarray}
The transformation \p{auxtr}-\p{auxtr1} has very important property.
Its superdeterminant (superJacobian)
\begin{equation}
\frac{\partial \{\eta'^m,y'^\mu\}}{\partial \{\eta^n,y^\nu\}} =
Ber^{-1}\left(\frac{\partial x'^M}{\partial x^N}\right)
\end{equation}
is inverse to the superJacobian for the transformation of the ``old"
coordinates $x^M$. This fact is based on the following property:
the superdeterminant of the graded matrix is inverse to the 
superdeterminant of the same matrix with opposite grading.
It means the invariance of the product 
$dV=d^Dx\underline{d}^{D_G}\theta d^{D_G}y\underline{d}^D\eta$
with respect to the general
coordinate reparametrization in the superspace. In turn this fact leads
to the invariance of the following integral
\begin{equation}\label{i}
I=\int dV F(x,\theta,\eta,y).
\end{equation}
If the integral over $\eta$ and $y$ exists, the result of such integration
\begin{equation}
I=\int d^Dx\underline{d}^{D_G}\theta\tilde{F}(x,\theta)
\end{equation}
will be invariant as well.

It is convenient at this stage to introduce additional set of auxiliary
variables $C_A$, transforming as the vector of the $GL(D,D_G)$ and
having the grassmann grading $g(C_a)=1,\;g(C_\alpha=0)$. With the
help of this ghosts it is easy to construct invariants in a covariant
manner under $SL(D,D_G)$. The additional volume element $dC=
\underline{d}C^DdC^{D_G}$ does not transforms due to condition $Ber h=1$
for $h$ belonging to $SL(D,D_G)$.

The simplest invariant has the form
\begin{equation}\label{ne}
I_0=\int dV dC e^{iE^AC_A}=\int dV dC e^{i\eta^M{E_M}^AC_A}.
\end{equation}
As was shown in \cite{N} the result of integration over $\eta^M$ and
$C_A$ in \p{ne} is proportional to the superdeterminant of the ${E_M}^A$
\begin{equation}
I_0=\int d^Dx\underline{d}^{D_G}\theta Ber {E_M}^A.
\end{equation}
The wide class of invariants can be obtained from the expressions of the
form
\begin{equation}\label{I}
I=\int dV dC F(\Omega, C)e^{iE^AC_A},
\end{equation}
where $F(\Omega, C)$ is an arbitrary function of $C_A$ and Cartan's
$\Omega$ -forms, in which differentials of coordinates $dx^M$ are changed
to $\eta^M$. Due to completeness of the one-form vielbein, the function 
$F(\Omega, C)$ can be represented as the series in powers of vielbein
and $C_A$ 
\begin{equation}
F(\Omega, C)=\sum {F^{B_k...B_1}}_{A_n...A_1}
E^{A_1}...E^{A_n} C_{B_1}...C_{B_k}
\end{equation}
with coefficients ${F_{B_k...B_1}}^{A_n...A_1}$ depending on the 
functions ${u^M}_{M_1...M_n}(x,\theta)$. 
Thus, the evaluation of the integral \p{I} reduces
to the evaluation of the basic integrals
\begin{equation}\label{basic}
{I^{A_1...A_n}}_{B_1...B_k}=\int d\eta dC 
E^{A_1}...E^{A_n} C_{B_1}...C_{B_k} e^{iE^AC_A}.
\end{equation}
One can show that such integrals are zero when numbers $n$ and $k$ are
different. The nonzero answers for two simplest cases are
\begin{eqnarray}
\int d\eta dC E^{A}C_{B} e^{iE^AC_A}&=& (-1)^A{\delta^A}_B Ber ({E_M}^C),\\
\label{res}
\int d\eta dC E^{A_1}E^{A_2}C_{B_1}C_{B_2} e^{iE^AC_A}&=&  \\
&&\!\!\!\!\!\!\!\!\!\!\!\!\!\!\!\!
\{(-1)^{A_1+A_2}{\delta^{A_2}}_{B_1}{\delta^{A_1}}_{B_2}+
(-1)^{A_1A_2+1}{\delta^{A_2}}_{B_2}{\delta^{A_1}}_{B_1}\}Ber {E_M}^A.
\nonumber
\end{eqnarray}
The general expression for the basic integrals \p{basic} can be evaluated
from the relation
\begin{equation}
\int d\eta dC e^{iE^A{\Sigma_A}^BC_B}=Ber ({E_M}^A{\Sigma_A}^B)=
Ber {E_M}^A Ber {\Sigma_D}^B
\end{equation}
by varying in it ${\Sigma_A}^B$ in the neighborhood of the identity.
The resulting ${I^{A_1...A_n}}_{B_1...B_k}$ look like 
expression \p{res} with correspondingly
symmetrized production of appropriate number of $\delta^{A_i}_{B_k}$
multiplied by $Ber {E_M}^A.$ This gives the answer for the invariant
$I$ \p{I} in terms of the superspace integral
\begin{equation}
I=\int d^Dx\underline{d}^{D_G}\theta\sum {F^{B_k...B_1}}_{A_n...A_1}
{I^{A_1...A_n}}_{B_1...B_k}.
\end{equation}

The question of finding the integral invariant of such type in the
superspace for the action of 
supergravity, like \p{act} for gravity, is open. If such action
does exist, it contains among the structure constants the Dirac
gamma matrices, $\gamma^a_{\alpha\beta}$, which break the tangent
space gauge group $GL(D)\times GL(D_G)$ down to its subgroup
(SL(2) in the case of four dimensions) and establish the connection
between bosonic and fermionic dimensionalities $D$ and $D_G$.

\setcounter{equation}0\section{Conclusions}
We have considered the nonlinear realizations of infinite - dimensional
diffeomorphism groups of any (super)space. The parameters of coset space
in a very natural manner include the coordinates,
 vielbeins and connections of the corresponding (super)space.
The geometrical and physical meaning of higher parameters 
${u^M}_{M_1...M_n}$ with $n\geq3$ is still unclear. Construction of 
invariant under the action of diffeomorphism group
differential $\Omega$ -forms is straightforward in any (super)space.
At the same time the $GL(D,D_G)$ gauge group, considered as the right
action on the group element, plays the role of gauge group in the
tangent space. The most of the $\Omega$-forms transform as
tensors of this $GL(D,D_G)$. The only $\Omega$-form with two tangent
indices $\Omega^a_b$ plays the role of connection which automatically is
torsionless in the bosonic case. In the superspace only 
 $T^a_{bc}$ and $T^\alpha_{\beta\gamma}$ components of the torsion
are vanishing identically.

Such an invariant differential $\Omega$ - forms can be considered as
building blocks for construction of integral invariants like action.
In purely bosonic space \p{act} gives the expression for the gravity
action. In the case of superspase the method of constructing integral
invariants is described in the Chapter 5. The existing of the action
for supergravity of such type is an open question.\vspace{0.5cm}\\

\noindent {\bf Acknowledgments.} I would like to stress the invaluable 
influence on me of  D.V. Volkov and  V.I. Ogievetsky. I would like to 
thank M.A. Vasiliev and A. Nersessian for useful discussions 
and comments on the subject.

This investigation has been supported in part by the
Russian Foundation of Fundamental Research,
grant 96-02-17634, joint grant RFFR-DFG 96-02-00186G,
and INTAS, grant 93-633, grant 94-2317 and grant of the
Dutch NWO organization.


\begin{thebibliography}{99}
\bibitem{BO}A.B. Borisov, V.I. Ogievetsky. Theor.Math.Phys., {\bf 21}
(1974) 329
\bibitem{OP}V.I. Ogievetsky. Lett.Nuovo Cim., {\bf 8} (1973) 988
\bibitem{IN}E.A. Ivanov, I. Niederle. Phys.Rev., {bf D45} (1992) 4545
\bibitem{W}J. Wess, J. Bagger. Supersymmetry and supergravity.
Princeton Univ.Press, 1983
\bibitem{B}F.A. Berezin. Introduction to the algebra and analysis with
anticommuting variables. Moscow Univ., 1983
\bibitem{BL}I.N. Bernstein, D.A. Leites. Functional Analysis, {\bf 11}
(1976) 70
\bibitem{N}P. van Nieuwenhuizen, Phys.Rep. {\bf C68} (1981) 189
\end{thebibliography}
\end{document}